\documentclass[twocolumn,aps,showpacs,superscriptaddress,pre]{revtex4}

\usepackage{amsmath,amssymb,graphicx}
\usepackage{color}
\usepackage{times}

\definecolor{yblue}{rgb}{0.06, 0.3, 0.57}
\usepackage[pdftex]{hyperref}
\hypersetup{colorlinks=true,linkcolor=blue,citecolor=blue,urlcolor=blue}

\DeclareRobustCommand{\ya}{\reflectbox{R}}

\begin{document}

\title{Patch-planting spin-glass solution for benchmarking}

\author{Wenlong Wang}
\affiliation{Department of Physics and Astronomy, Texas A\&M University, 
College Station, Texas 77843-4242, USA}

\author{Salvatore Mandr{\`a}}
\affiliation{NASA Ames Research Center Quantum Artificial Intelligence 
Laboratory (QuAIL), Mail Stop 269-1, Moffett Field, California 94035, USA}
\affiliation{Stinger Ghaffarian Technologies Inc., 7701 Greenbelt Rd., 
Suite 400, Greenbelt, Maryland 20770, USA}

\author{Helmut G. Katzgraber}
\affiliation{Department of Physics and Astronomy, Texas A\&M University, 
College Station, Texas 77843-4242, USA}
\affiliation{1QB Information Technologies (1QBit), Vancouver, British
Columbia, Canada V6B 4W4}
\affiliation{Santa Fe Institute, 1399 Hyde Park Road, Santa Fe, 
New Mexico 87501, USA}

\begin{abstract}

We introduce an algorithm to generate (not solve) spin-glass instances
with planted solutions of arbitrary size and structure. First, a set of
small problem patches with open boundaries is solved either exactly or
with a heuristic, and then the individual patches are stitched together
to create a large problem with a known planted solution. Because in
these problems frustration is typically smaller than in random problems,
we first assess the typical computational complexity of the individual
patches using population annealing Monte Carlo, and introduce an
approach that allows one to fine-tune the typical computational
complexity of the patch-planted system. The scaling of the typical
computational complexity of these planted instances with various numbers
of patches and patch sizes is investigated and compared to random
instances.

\end{abstract}

\maketitle

Many optimization problems belong to the NP-hard complexity class, for
which it is believed that no algorithms exist to solve them in
polynomial time. Spin-glass problems without biases and on nonplanar
topologies, such as the Edward-Anderson (EA) model \cite{edwards:75},
represent a sub-class of the NP-hard class.  Because spin glasses are
the simplest models with both disorder and frustration that fall into
the NP-hard class, they represent the ideal model systems to benchmark
algorithms, as well as novel computing architectures.  A number of
heuristics, as well as exhaustive search methods, have been designed and
developed to minimize spin-glass Hamiltonians as efficiently as
possible.  These method also include simulated annealing
\cite{kirkpatrick:83}, parallel tempering Monte Carlo
\cite{swendsen:86,geyer:91,hukushima:96,moreno:03}, population annealing
Monte Carlo \cite{hukushima:03,zhou:10,wang:15,wang:15e}, and genetic
algorithms \cite{pal:96,hartmann:01}, as well as
branch-and-cut \cite{desimone:95} algorithms, to name a few. Many of
these optimization algorithms use only local updates during the
minimization procedure. However, in many cases, the use of cluster
algorithms with nonlocal updates can greatly enhance the searching
process when the energy landscape has many metastable states with small
overlap \cite{houdayer:01,zhu:15b,zhu:16y}.  In the last two decades,
quantum heuristics have been proposed as an alternative to classical
heuristics, due to their potential to exploit quantum
superposition and quantum tunneling effects. Among quantum approaches,
adiabatic quantum optimization (AQO) is widely used
\cite{kadowaki:98,nishimori:01,farhi:01,santoro:02,roland:02,boixo:13a,boixo:14,ronnow:14a,boixo:16,pudenz:15,perdomo:15,vinci:15,mandra:15,venturelli:15a}
and likely the method most amenable to hardware implementations
\cite{johnson:11}. Current state-of-the-art AQO hardware is manufactured
by D-Wave System Inc., whose latest chip allows for the quantum
optimization of problems of approximately up to $2000$ variables.
However, whether AQO can be more efficient than classical algorithms for
certain problems is still controversial
\cite{heim:15,mandra:16b,mandra:17}.

\begin{figure}[htb]
\includegraphics[width=1.0\columnwidth]{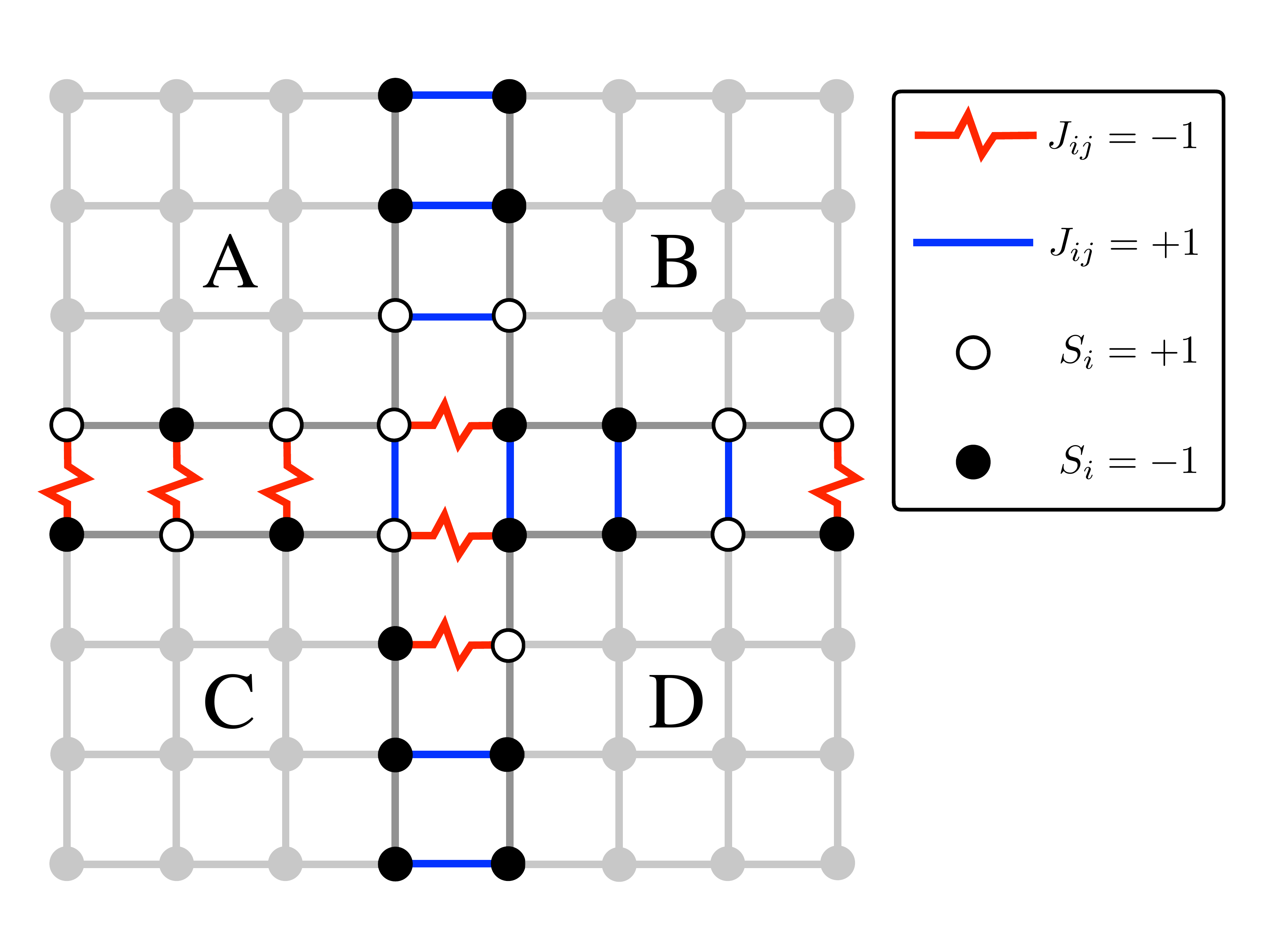}
\vspace*{-3.0em}
\caption{
Schematic diagram of the patch planting method for a two-dimensional
lattice. First, the ground state for each (easily solvable) patch (A -- D)
is computed using free boundary conditions. Then, a ground state
configuration of each patch is selected. Edge spins between the patches
are represented by empty and full circles.  Couplers between the edge
spins between adjacent patches are added under the condition that all
interactions are satisfied, i.e., if two spins have the same value
(i.e., both full or both empty circles), a ferromagnetic coupler
(straight blue line) is added. If, however, two adjacent spins are
different, then an antiferromagnetic coupler (red wiggly line) is added.
The direct product of the ground state of the patches A -- D is then
the ground state of the large planted system.
}
\label{PP}
\end{figure}

Given the importance of comparing optimization techniques across
disciplines, it is necessary to have benchmark problems that are (1)
representative of the hardness of a typical NP-hard problem, (2)
scalable for large systems, and for which (3) the ground state is
known {\em a priori}. While it is easy to fulfill criteria (1) and (2),
it is challenging to have large problems with known solutions. 

There have been previous approaches to plant solutions for benchmarking
purposes. For example, Ref.~\cite{hen:15a} used an approach based on
constraint satisfaction problems. Although these problems are tunable in
hardness, there is little control when selecting the coupler values
between the individual variables. For analog machines with finite
precision, such as the D-Wave quantum annealers, this could be an
unnecessary restriction.  Other approaches \cite{marshall:16} start from
a random coupler configuration and then stochastically update the values
of the couplers with a penalty that directly correlates to the
time-to-solution of a given solver. However, this approach has two
shortcomings: Fist, it assumes that the typical computational hardness
\cite{comment:typical} of a problem for a given algorithm will carry
over to other optimization techniques. Second, for extremely large
problems, the stochastic approach will take sizable resources to
thermalize and thus will not be practical.

The method we propose here and which we call ``patch planting'' (see
Fig.~\ref{PP}), where we solve small problems (patches) with open
boundaries and then stitch these together to plant an arbitrarily large
solution to an instance, does not suffer from these shortcomings: First,
arbitrarily large problems can be generated. Second, by assessing the
typical complexity using the entropic family size of population
annealing Monte Carlo, a metric that characterizes the landscape of
the problem and not the algorithmic complexity, we do not depend on
the behavior of a particular algorithm when assessing the typical time
to solution for a particular instance.  Finally, the method poses no
restrictions to coupler values, biases (field terms), or lattice
topologies.

The paper is structured as follows. In Sec.~\ref{sec:model} we introduce
the benchmark problem, as well the patch planting algorithm.  In
Sec.~\ref{sec:res} we use simulated annealing, population annealing
Monte Carlo, as well as experiments on the D-Wave 2X quantum annealer to
illustrate how patch planting can produce computational problems that
are typically hard.  Concluding remarks are presented in
Sec.~\ref{sec:summary}.

\section{Patch planting}
\label{sec:model}

The patch planting heuristic can be described via the following steps:
\begin{itemize}

\item[(i)] Find the ground state of patches using \textit{free boundary
conditions}.

\item[(ii)] For each patch, choose an arbitrary ground-state
configuration.

\item[(iii)] Connect the patches with couplers between the free boundary
spins ensuring that all couplings are satisfied.

\end{itemize} Note that the patches can be chosen arbitrarily, as long
as they can be glued together to form the desired problem or topology with
the edges to be stitched together having free boundaries. In addition,
the individual patches can be solved with any available optimization
technique. As demonstrated below, it is important to solve as large
patches as possible, because this will result in problems of comparable
computational complexity to purely random problems.  In some cases, the
breakup of a problem might result in a patch that can be solved exactly,
i.e., in polynomial time. Finally, when stitching the patches together,
as shown in Fig.~\ref{PP}, it is important to ``satisfy'' the interaction
between two spins of different patches. This means that the coupler has
to be chosen as to minimize the dimer's energy.  Knowing the minimizing
configuration of the individual patches and assigning the stitching
couplers as to satisfy the interactions between spins of neighboring
patches then results in a larger planted solution \cite{comment:alan}.

As described in Sec.~\ref{sec:res} in more detail, the typical
computational complexity of the patched problem can be tuned by either
changing the patch size (the larger, the harder) or using hard patches
(the harder the patch, the harder the compound problem) e.g., by
measuring the entropic family size via population annealing Monte Carlo.
This metric can be measured with little numerical effort, and gives a
good representation of the typical computational complexity of a
problem. Therefore, by post-selecting individual patches, problems of
different typical computational complexity can be generated.

Note that in the description of the patch planting procedure no details
of the problem to be studied have been mentioned, because the
approach is agnostic to the choice of couplers and topologies.  We thus
emphasize that the patch planting approach can be used for problems of
{\em arbitrary} topology and for an arbitrary set of coupler values and
biases. As such, solutions for arbitrary problems can be planted. This
is of much importance when attempting to generate problem sets with
particular features, such as synthetic application problems that are
known to have a specific nonrandom structure, or problems where the
minimum energy gap is fixed (and large) to mitigate the effects of noise
on analog optimization machines \cite{katzgraber:15,zhu:16}.

\section{Experiments}
\label{sec:res}

\subsection{Benchmark problem}

To test the properties of patch planting, we use the Edward-Anderson
(EA) Ising spin-glass model \cite{edwards:75}, initially in three space
dimensions.  Later, we perform experiments on the D-Wave 2X quantum
annealer using the native topology of the machine \cite{bunyk:14}.  The
EA spin glass is defined by the following Hamiltonian
\begin{equation}
H = - \sum_{ij} J_{ij} S_i S_j -\sum_{i} h_i S_i \, ,
\label{eq:ham}
\end{equation}
where $S_i \in \{\pm 1\}$ are Ising spins and the first sum is over
spin-spin interactions. For a three-dimensional lattice, the sum is over
nearest neighbors on a cubic lattice.  For simplicity, all the local
magnetic fields are set to zero, i.e., $h_i = 0$.  We do emphasize,
however, that patch planting also works with external biases.  The
spin-spin interactions $J_{ij}$ are chosen from a normal distribution
with zero mean and unit variance.  A set of the couplings defines an
``instance.''

Given the hardware limitations of the D-Wave quantum chips, instances
for the D-Wave 2X have been created by planting and patching together
K$_{44}$ unit cells following the two-dimensional logical structure of
the Chimera graph.  The couplers are randomly drawn from the Sidon set
$\{\pm 5,\pm6,\pm7\}$ \cite{zhu:16}. In both cases, we use free boundary
conditions (FBCs) for the patches to plant larger instances. We also
compare our patched instances with free boundary conditions to random
instances with periodic boundary conditions (PBCs).

\begin{table}
\caption{
Simulation parameters for the three-dimensional EA model experiments
using population annealing Monte Carlo. Here, $L_0$ is the patch size,
$L$ is the linear system size, $R$ is the number of replicas used in the
simulation, $T_0 = 1/\beta_0$ is the lowest temperature simulated, $N_T$
is the number of temperature steps (evenly spaced in $\beta$) in the
annealing schedule, BC is the type of boundary condition [either
periodic boundary condition (PBC) or free boundary condition (FBC)],
and $N_{\rm sa}$ is the number of disorder realizations studied. For
each replica, $N_S=10$ Monte Carlo sweeps are performed at each
temperature during the anneal. Data for PBC with $L=8$ are taken from
Ref.~\cite{wang:15}.
\label{para}
}
{\scriptsize
\begin{tabular*}{\columnwidth}{@{\extracolsep{\fill}} l l l l l l r}
\hline
\hline
$L_0$ &$L$  & $R$  & $T_0$ & $N_T$ &BC & $N_{\rm sa}$ \\
\hline
4 &4  & $4\times10^3$ & $0.2$  & $101$ & FBC & $345600$ \\
4 &8  & $10^4$ & $0.2$  & $101$ & FBC & $5000$ \\
4 &12 & $5\times10^4$ & $0.2$  & $201$ & FBC & $5120$ \\
4 &16 & $2\times10^5$ & $0.2$  & $301$ & FBC & $1877$ \\
4 &20 & $10^6$ & $0.2$  & $401$ & FBC & $194$ \\
5 &5  & $10^4$ & $0.2$  & $101$ & FBC & $345600$ \\
5 &10 & $10^5$ & $0.2$  & $201$ & FBC & $5000$ \\
6 &6  & $2\times10^4$ & $0.2$  & $101$ & FBC & $41472$ \\
6 &12 & $10^5$ & $0.2$  & $201$ & FBC & $1752$ \\
8 &8  & $5\times10^4$ & $0.2$ & $201$ & FBC & $23358$ \\
8 &16 & $10^6$ & $0.2$  & $301$ & FBC & $624$ \\
10 &10 & $10^6$ & $0.2$  & $301$ & FBC & $8000$ \\
10 &20 & $2\times10^6$ & $0.2$  & $401$ & FBC & $260$ \\
\hline
8 &8   & $10^5$ & $0.2$ & $101$ & PBC & $5099$ \\
12 &12 & $10^6$ & $0.2$ & $201$ & PBC & $3812$ \\
\hline
\hline
\end{tabular*}
}
\end{table}

\subsection{Simulation details}

We use the entropic family size of population annealing Monte Carlo
$\rho_s$ \cite{wang:15e} to characterize the hardness of the instances.
All simulation parameters for the three-dimensional
Edwards-Anderson model are listed in Table~\ref{para}.  For the Chimera
graph studies on the D-Wave 2X machine, we find the ground state of the
patches using $R=2 \times 10^5$ population members, $N_T=301$
temperature steps, $N_S=10$ Monte Carlo sweeps, and $T_0=0.1$ the lowest
temperature of the anneal. The simulation for random problems
are done with the same parameters, except $R=10^6$.

Experiments on the D-Wave 2X quantum annealer have been performed using
a chip with $N = 1097$ working qubits.  For the Chimera graph, we used
all available qubits and patched the system using either two, three, or
four patches, respectively. That is, if the system has $12 \times
12$ K$_{44}$ cells of eight qubits each, we divide the whole lattice into
two patches of $6 \times 12$ K$_{44}$ cells, three patches of $4 \times 12$
K$_{44}$ cells, or four patches of $3 \times 12$ K$_{44}$ cells.  For the
experiments, we used an annealing time of $20\ \mu$s, $100$ gauges, and
$1000$ readouts for each gauge.

\subsection{Correlation between typical hardness and the entropic
family size}

The first crucial step in investigating the hardness of instances is to
find a good metric that reliably characterizes the typical problem
complexity, yet is easy to measure with little computational cost.  One
approach would be to use the success probability of simulated annealing
as a proxy.  However, even for medium-size systems, this metric becomes
unreliable and computationally expensive. Another possibility consists
in using specialized classical algorithms \cite{marshall:16}, such as
the Hamze--de Freitas--Selby heuristic \cite{hamze:04,selby:14}.
However, in this case the typical computational complexity depends on a
chosen algorithm and not on the intrinsic properties of the problem's
energy landscape. The latter can be mapped out well for random problems
using parallel tempering Monte Carlo \cite{katzgraber:15}, however, at
sizable computational cost for large patches.  Therefore, in this work we
infer the typical hardness of instances through the entropic family size
$\rho_s$ of population annealing Monte Carlo.

\begin{figure}[htb]
\includegraphics[width=1.0\columnwidth]{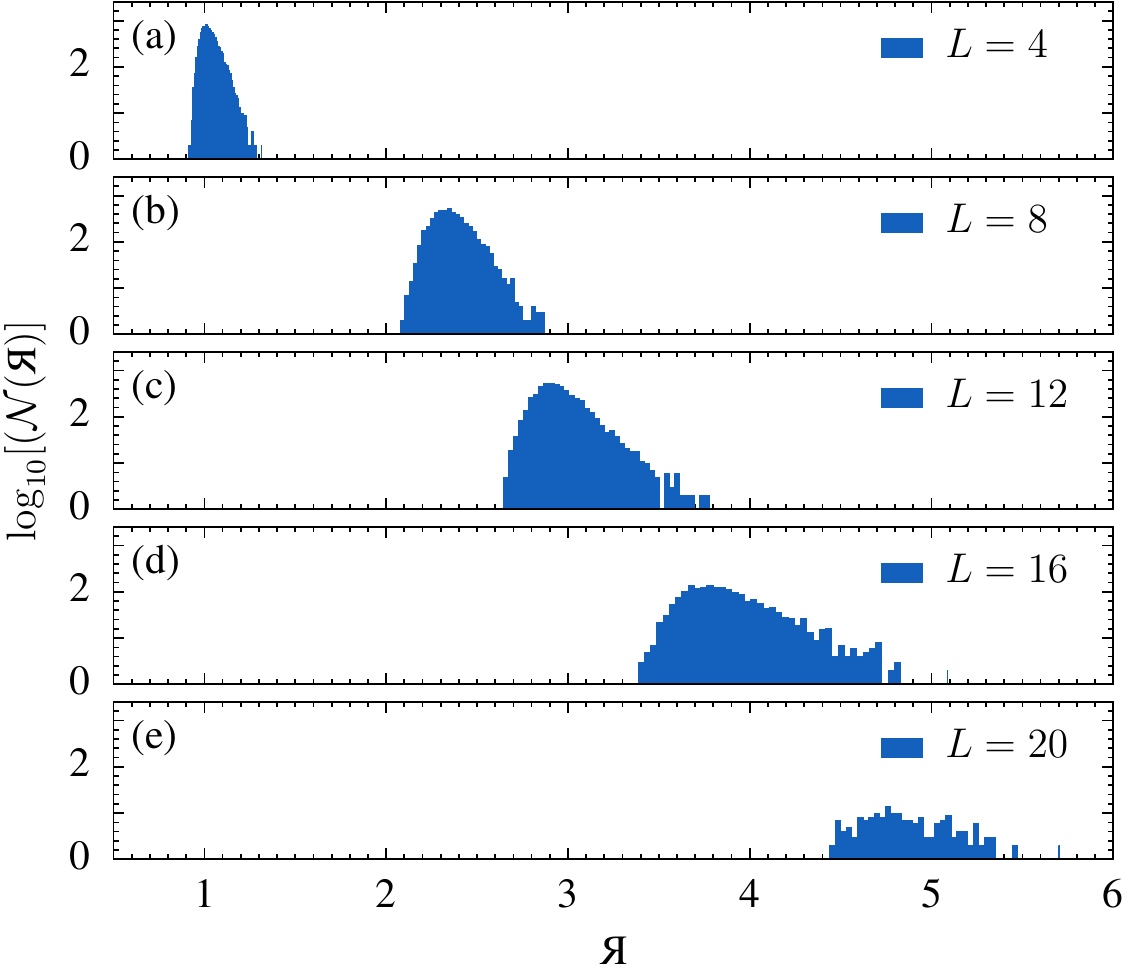}
\caption{ 
Distribution of $\ya$; [defined in Eq.~\eqref{eq:ya}] for various system sizes
[$L=4$ (a), $8$ (b), $12$ (c), $16$ (d), and $20$ (e)] in three space
dimensions. For increasing system size, the typical complexity of the
problems grows, which is mirrored by the distributions of $\ya$ shifting
to the right.
}
\label{L04}
\end{figure}

\begin{figure}[htb]
\hspace*{-2.0em}\includegraphics[width=0.8\columnwidth]{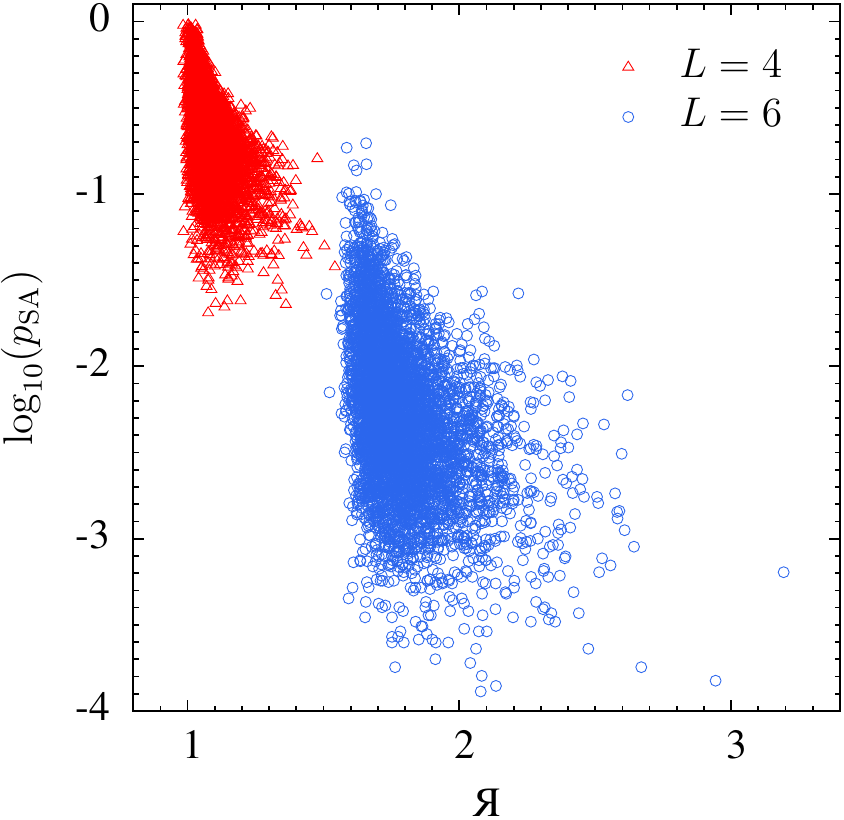}
\caption{
Correlation of the probability to find the ground state of simulated annealing
$p_{\rm{SA}}$ and the log of entropic family size of population
annealing $\ya$ for three-dimensional systems with $L=4$ and $L=6$ at
$\beta=5$ (data taken from Refs.~\cite{wang:14} and \cite{wang:15}).
When $\ya$ is larger, it is also more difficult to find the ground
state, i.e., $p_{\rm{SA}}$ is smaller. Note that $p_{\rm{SA}}$ drops
very rapidly as $L$ increases, while it is easier to measure
$\ya$.}
\label{PR}
\end{figure}

Population annealing (PA) Monte Carlo \cite{hukushima:03,machta:10} is
closely related to simulated annealing (SA), except that it uses a
population of $R$ replicas and the population is resampled at each
temperature anneal step to maintain thermal equilibrium.  At each
simulation step, replicas are duplicated accordingly to the ratio
between the Boltzmann factors computed after and before updating the
temperature. This means that replicas with lower (higher) energy tend to
be duplicated (eliminated), ensuring the correct representation of the
Boltzmann distribution. Therefore, PA improves the probability to find
the lowest energy state over SA by more efficiently sampling phase
space. We choose to normalize our replicas so that the population size
stays approximately the same. Similar to SA, Metropolis sweeps are
applied to each replica at the new temperature.  At low temperatures,
most of the original population is eliminated in the resampling steps
and the final population is a descendant from a small subset of the
initial population. Let $n_i$ be the fraction of the population from
family $i$ in the initial population, then
\begin{equation}
\rho_s=\lim\limits_{R \rightarrow \infty} R \times e^{\sum\limits_i n_i \log
n_i}\,. 
\end{equation}
Here, $\rho_s$ represents the characteristic survival family size. The
larger $\rho_s$ is, the less surviving families, i.e., the more rugged
the energy landscape. Moreover, $\rho_s$ correlates strongly with the
integrated autocorrelation time of parallel tempering, which is also a
proxy towards the roughness of the energy landscape \cite{wang:15e}.
Note that $\rho_s$ converges quickly in population size and is easily
estimated with simulations. See Ref.~\cite{wang:15e} for more details on
population annealing.  Because $\rho_s$ is approximately log-normal
distributed (see Fig.~\ref{L04}), let us define the logarithm of $\rho_s$ as
\begin{equation}
\ya=\log_{10}(\rho_s)\,.
\label{eq:ya}
\end{equation}
Figure~\ref{PR} shows the correlation between the probability to find
the ground state for SA, $p_{\rm SA}$ at inverse temperature $\beta =
1/T = 5$ and $\ya$ (data taken from Refs.~\cite{wang:14,wang:15}). As
expected, the probability of success decreases by increasing $\ya$.
Indeed, SA struggles more to find the ground state when the energy
landscape is more rugged. Therefore, $\ya$ represents a good metric to
estimate the typical hardness of optimization problems. In this work,
because we study large patched system sizes in three dimensions, we have
used $\ya$ at $\beta=3$, which is still a low temperature compared to
the spin-glass transition temperature for this model
\cite{katzgraber:06}. For the Chimera graph, where there is no phase
transition, we have used $\ya$ at a considerably lower temperature
$\beta=10$.

\subsection{Results in three space dimensions}

We first focus on the scaling properties of $\ya$ for patch-planted
instances by either varying the patch sizes $L_0$ or the system size
$L$. In addition, we also demonstrate that harder patches can be used to
patch harder instances.

\begin{figure}[htb]
\includegraphics[width=1.0\columnwidth]{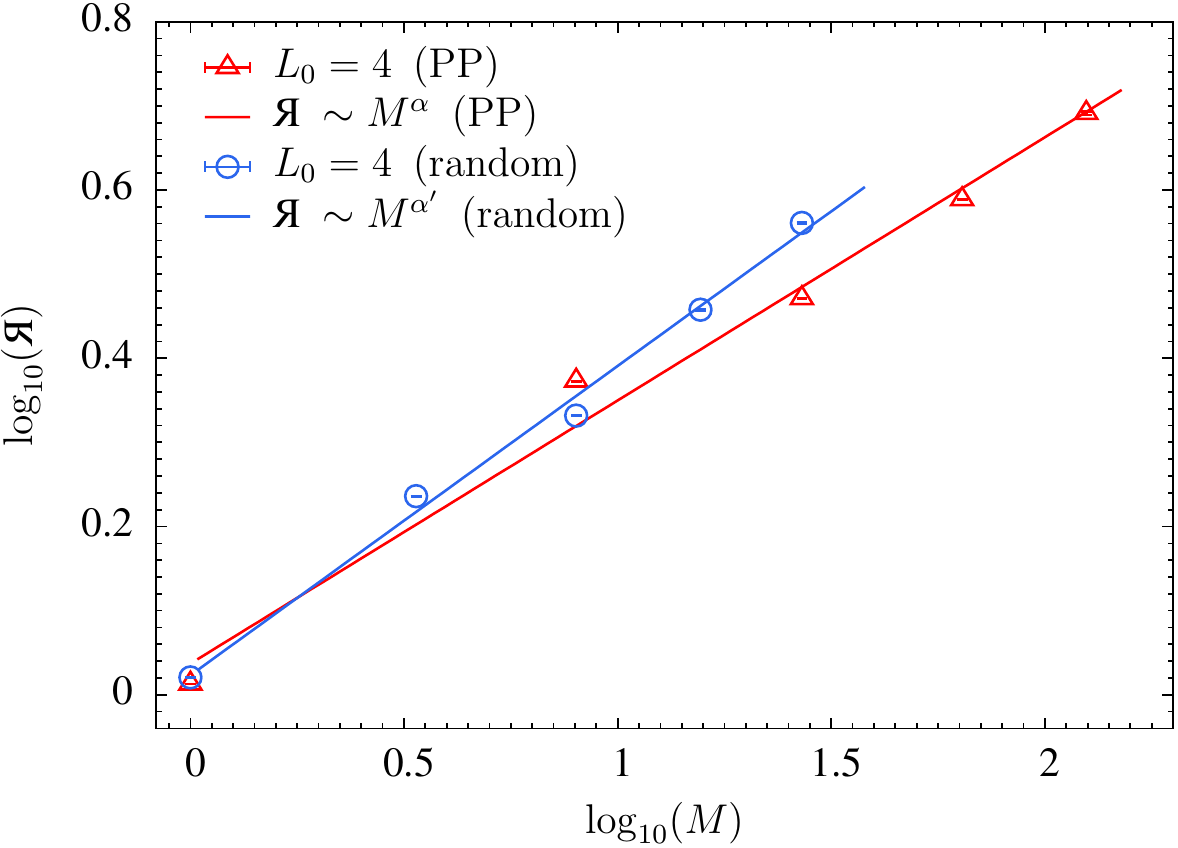}
\caption{
Scaling of the logarithm of the entropic family size $\ya =
\log_{10}\rho_s$ by varying the number of patches $M$ (triangles,
labeled with ``PP''). The line is a power-law fit of the form $\ya(L_0)
M^\alpha$, where $\ya(L_0)$ is $\ya$ of a single patch. From the fit, we
obtain $\alpha = 0.31(3)$. We also compare to random problems on a
three-dimensional lattice (circles). In this case a power-law fit
results in $\alpha' = 0.37(4)$, i.e., the two classes scale similarly,
yet with two different exponents.}
\label{L04rho}
\end{figure}

\begin{figure}[htb]
\includegraphics[width=1.0\columnwidth]{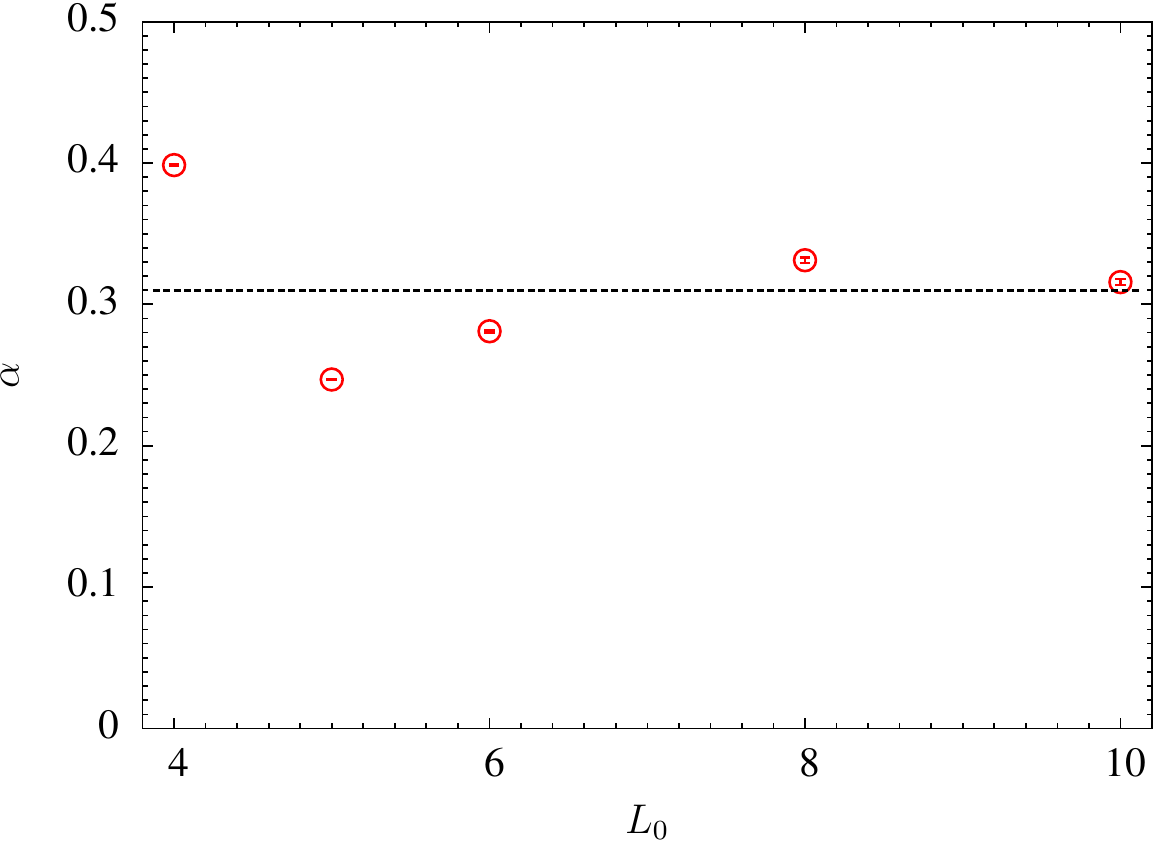}
\caption{
Scaling exponent $\alpha$ ($\ya \sim M^\alpha$; see Fig.~\ref{L04rho}),
by varying the patch size $L_0$ but keeping the number of
patches fixed to $M = 8$.}
\label{alpha}
\end{figure}

Let $M=(L/L_0)^3$ the number of patches of size $L_0$.  For random
instances, $\rho_s$ grows exponentially with $L$ \cite{wang:15e}.
Because one would expect that $\rho_s$ for a problem of size $L$ by
patching $M$ patches of size $L_0$ cannot be larger than the product of
$\rho_s$ of the individual patches, the patched instance complexity is
bounded,
\begin{equation}\label{eq:boundXi}
  \ya(M, L_0) \leq M \ya(L_0),
\end{equation}
where $\ya(M, L_0)$ is $\ya$ of the patched instance of $M$ patches of
size $L_0$ and $\ya(1, L_0) \equiv \ya(L_0)$ is $\ya$ of a patch.  In
Fig.~\ref{L04rho}, we show the scaling of $\ya$ by varying the number of
patches $M$ and a power-law fit of the form
\begin{equation}\label{eq:scalingXi}
  \ya(M,L_0) = \ya(L_0) M^\alpha,
\end{equation}
where $0 < \alpha < 1$.  $\ya$ scales sub-linearly with $M$ with an
exponent $\alpha = 0.31(3)$. This proves that the patch planted
instances become \emph{harder} by increasing the system size via the
number of patches.  Therefore, it is guaranteed that, for a sufficiently
large number of patches, patch planted instances can become arbitrarily
hard in the thermodynamic limit. Figure \ref{alpha} shows the scaling of
the exponent $\alpha$ by increasing the size of the patches $L_0$, while
keeping the number of patches fixed to $M = 8$.  As one can see from the
figure, $\alpha$ remains roughly constant for a wide range of $L_0$
values implying that $\alpha$ is a characteristic constant for patch
planted problems.  It is interesting to compare the scaling with random
instances by defining an effective number of blocks as $M= (L/L_0)^3$,
also shown in Fig.~\ref{L04rho}. We find that both random and patch
planted instances have a similar scaling form, although the random class
has a larger exponent $\alpha'=0.37(4)$, as expected. Therefore,
$\rho_s$ for patched instances also approximately scales exponentially
with system size $L$, as is the case for random instances. Note that
$\alpha$ and $\alpha'$ likely depend on the characteristics of the
problem to be studied.

\begin{figure}[htb]
\includegraphics[width=1.0\columnwidth]{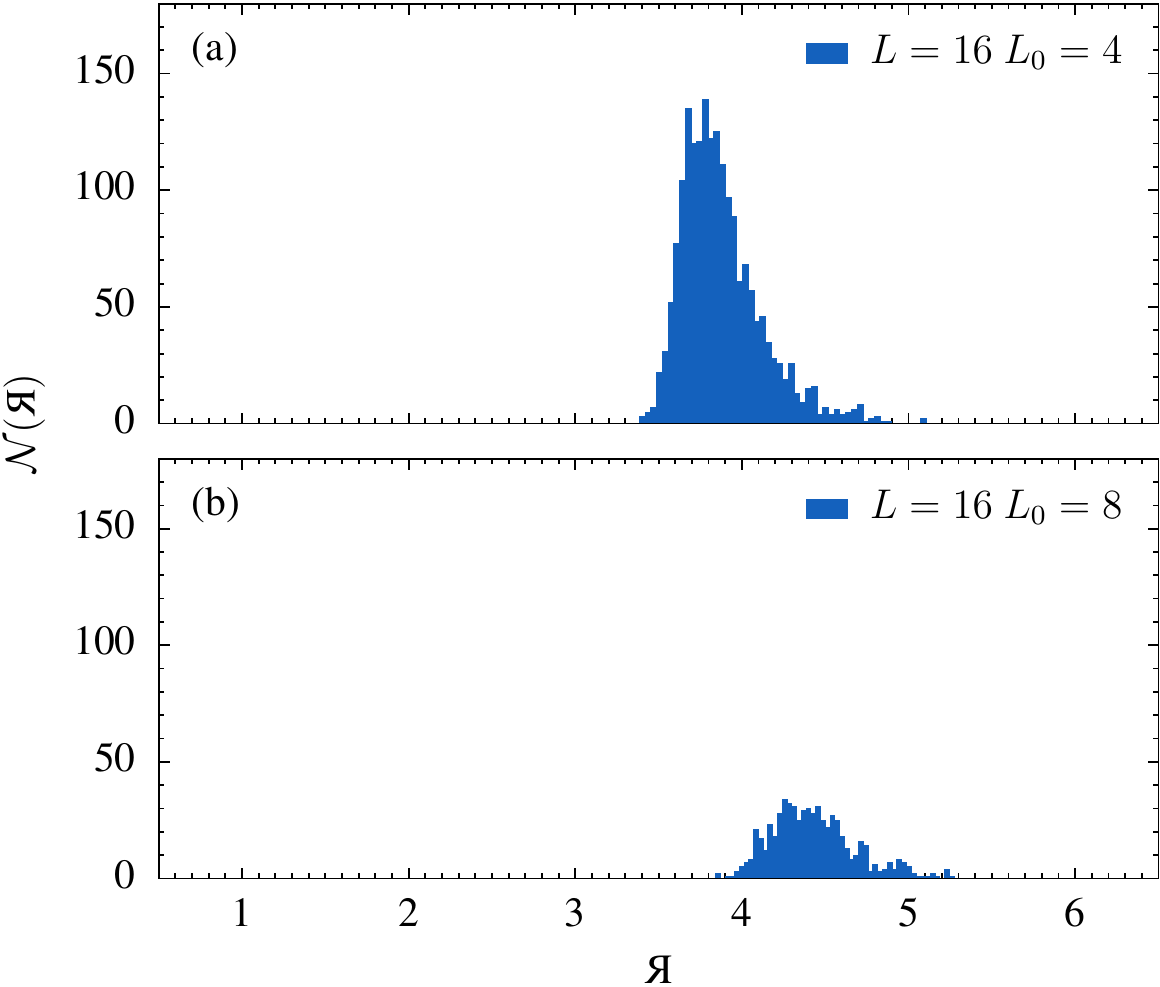}
\caption{
Comparison of the distribution of $\ya$ for $L=16$ in three space
dimensions, but with different patch sizes $L_0=4$ (a) and $8$ (b). 
There is a noticeable shift in the distributions of $\ya$.
Therefore, to patch harder instances, one should use as large patch
sizes as possible.
}
\label{Block48}
\end{figure}

\begin{figure}[htb]
\includegraphics[width=1.0\columnwidth]{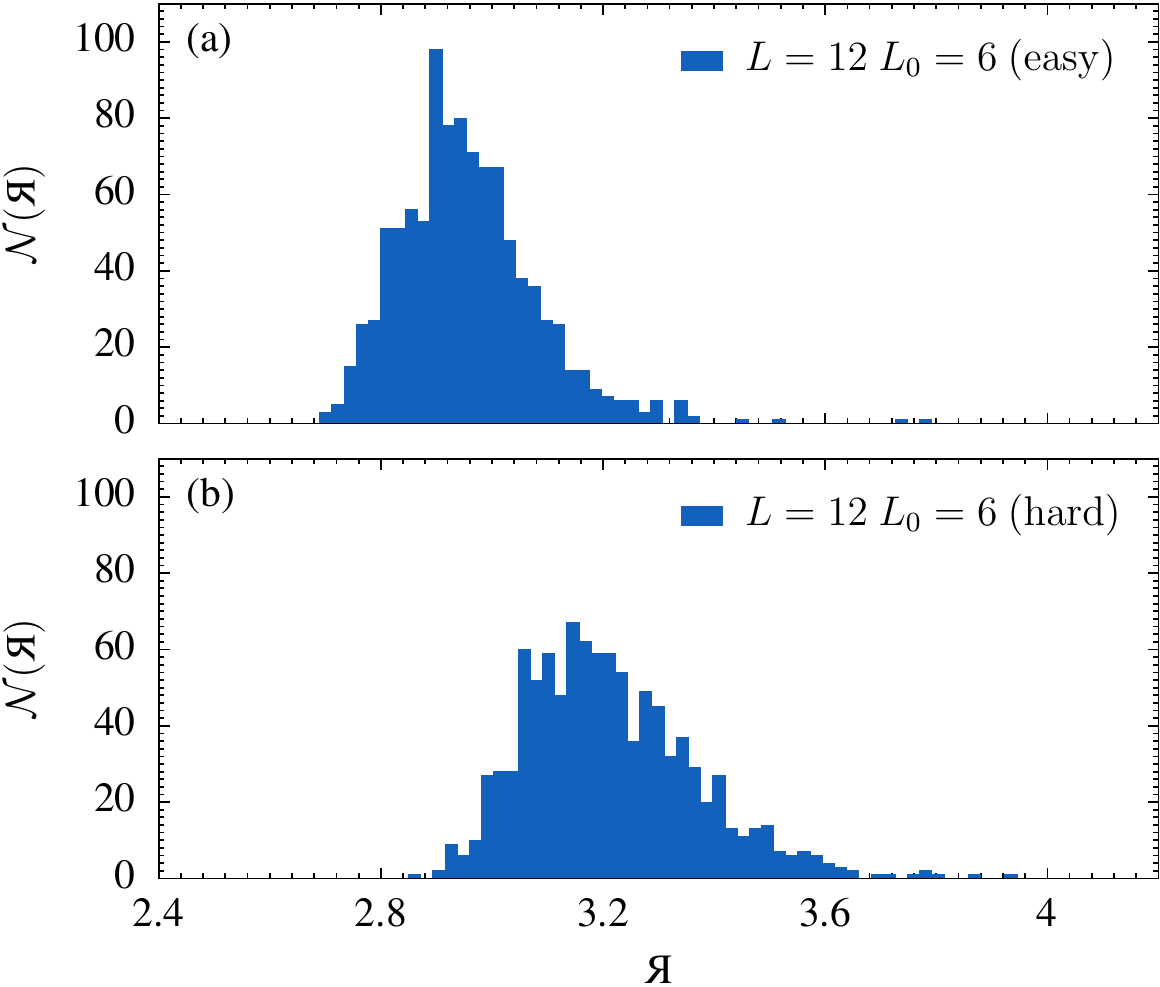}
\caption{
Distributions of $\ya$ for $M = 8$ ($L = 12$) patches of size $L_0=6$
using either easy patches (a) or hard patches (b). As
expected, the distribution shifts to the right when harder patches
are used.
}
\label{Xi12}
\end{figure}

\begin{figure}[htb]
\includegraphics[width=1.0\columnwidth]{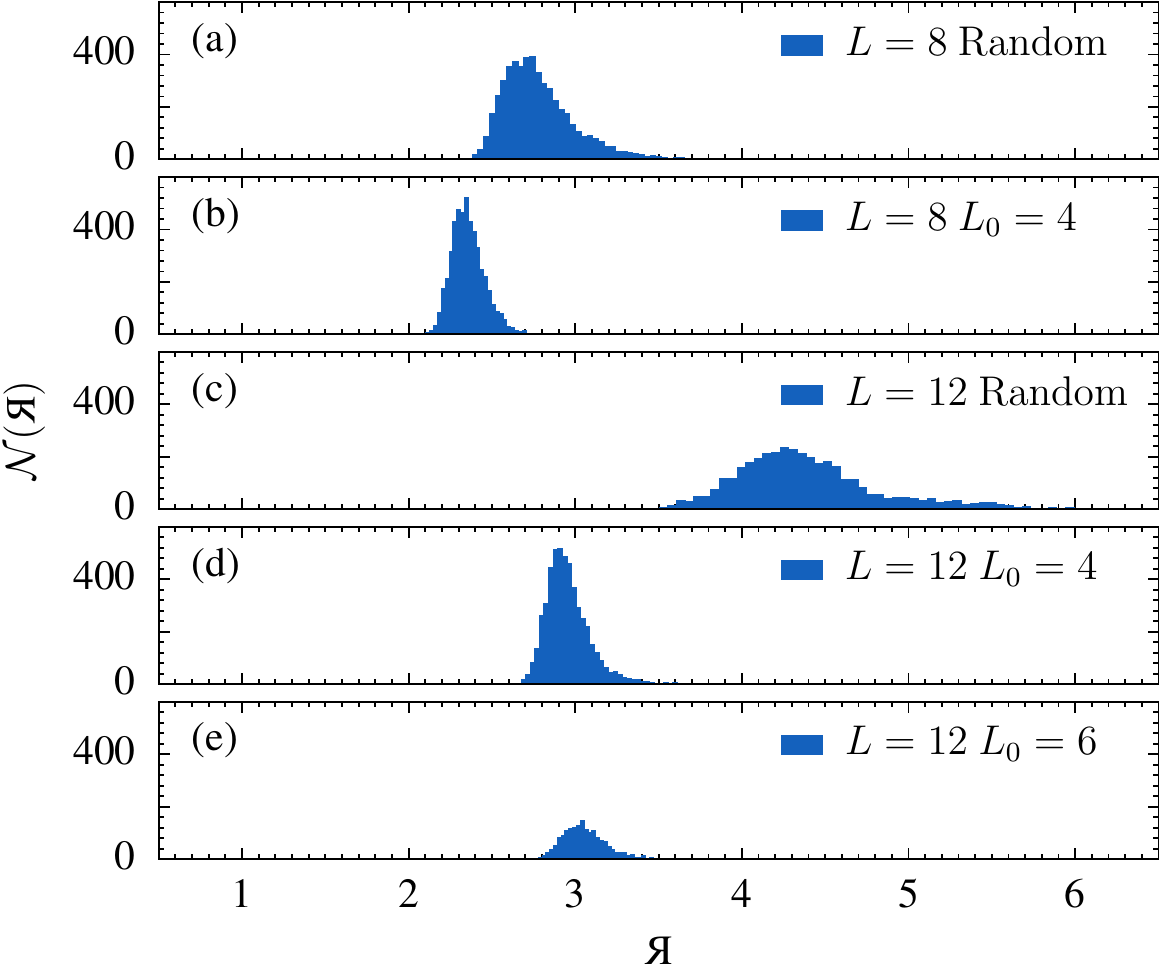}
\caption{
Comparison of the typical complexity of the patched instances with
random instances. The patched instances shown in panels (b), (d), and
(e) are generally computationally easier than their random counterparts
shown in panels (a) and (c), for a given system size. Note the overlap
between the distributions, i.e., by mining the data one can obtain very
hard patched problems.
}
\label{RH}
\end{figure}

One may also expect to have some benefit by using either larger or
harder patches. Indeed, in both cases, this results in having a larger
value of $\ya$. In Fig.~\ref{Block48} we show the effects of having
larger patches by analyzing the distribution of $\ya$ at fixed size of
the system, $L = 16$, using two different patch sizes, $L_0 = 4$
[Fig.~\ref{Block48}(a)] and $L_0 = 8$ [Fig.~\ref{Block48}(b)]. As one
can see, patched instances are consistently harder by using larger
patches for a fixed system size.  Similarly, in Fig.~\ref{Xi12} we show
the distribution of $\ya$ by patching instances with $M = 8$ patches of
size $L_0 = 6$ by either using {\em easy} [Fig.~\ref{Xi12}(a)] or {\em
hard} [Fig.~\ref{Xi12}(b)] patches. We defined {\em easy} patches as the
$8000$ patches with the smallest $\ya$ and hard patches as the $8000$
patches with the largest $\ya$ from the $41472$ patches randomly
generated. From these, $1000$ easy and $1000$ hard instances are then
generated.

\begin{figure}[htb]
\includegraphics[width=0.47\columnwidth]{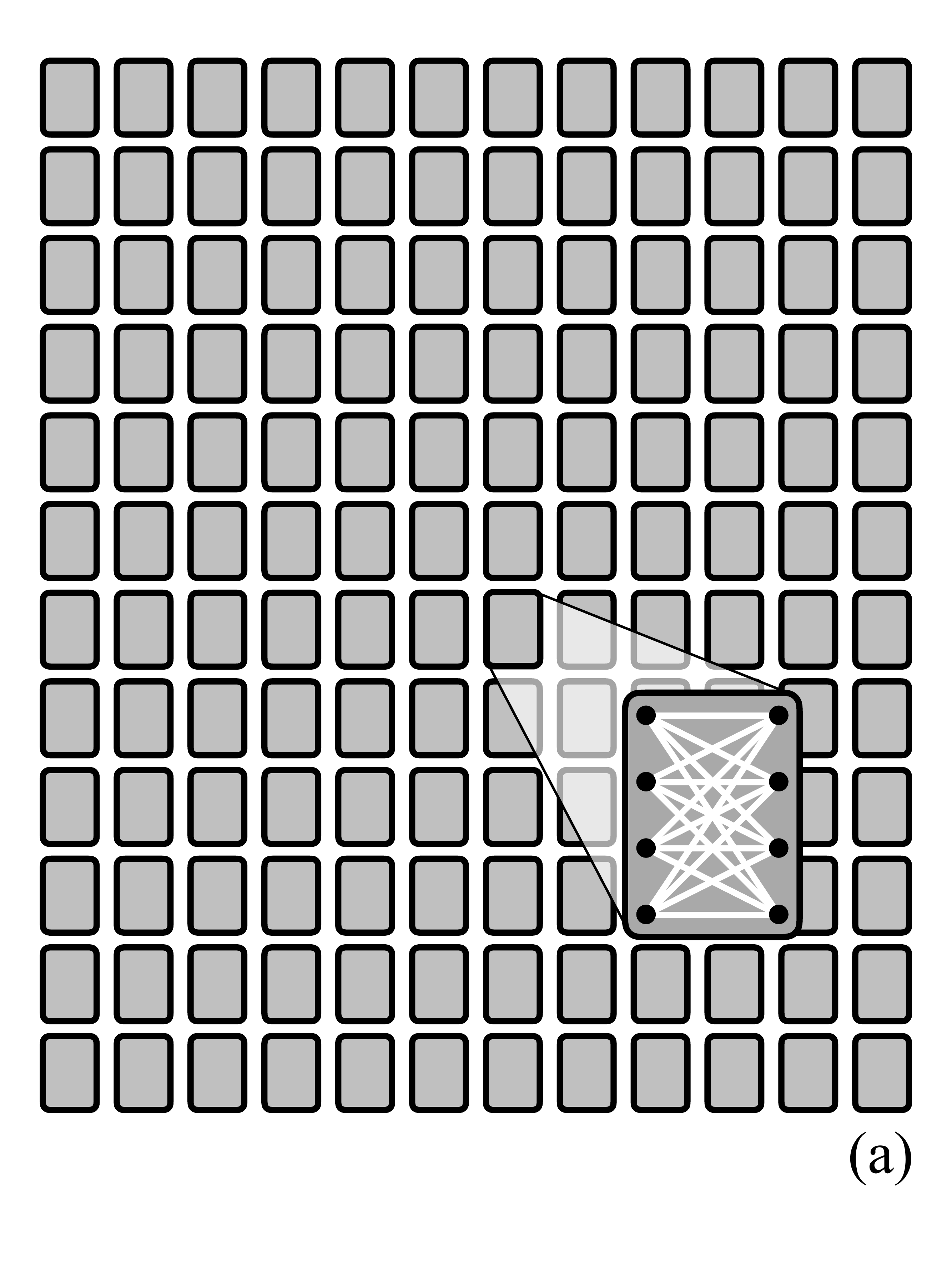}
\hspace*{0.25em}
\includegraphics[width=0.47\columnwidth]{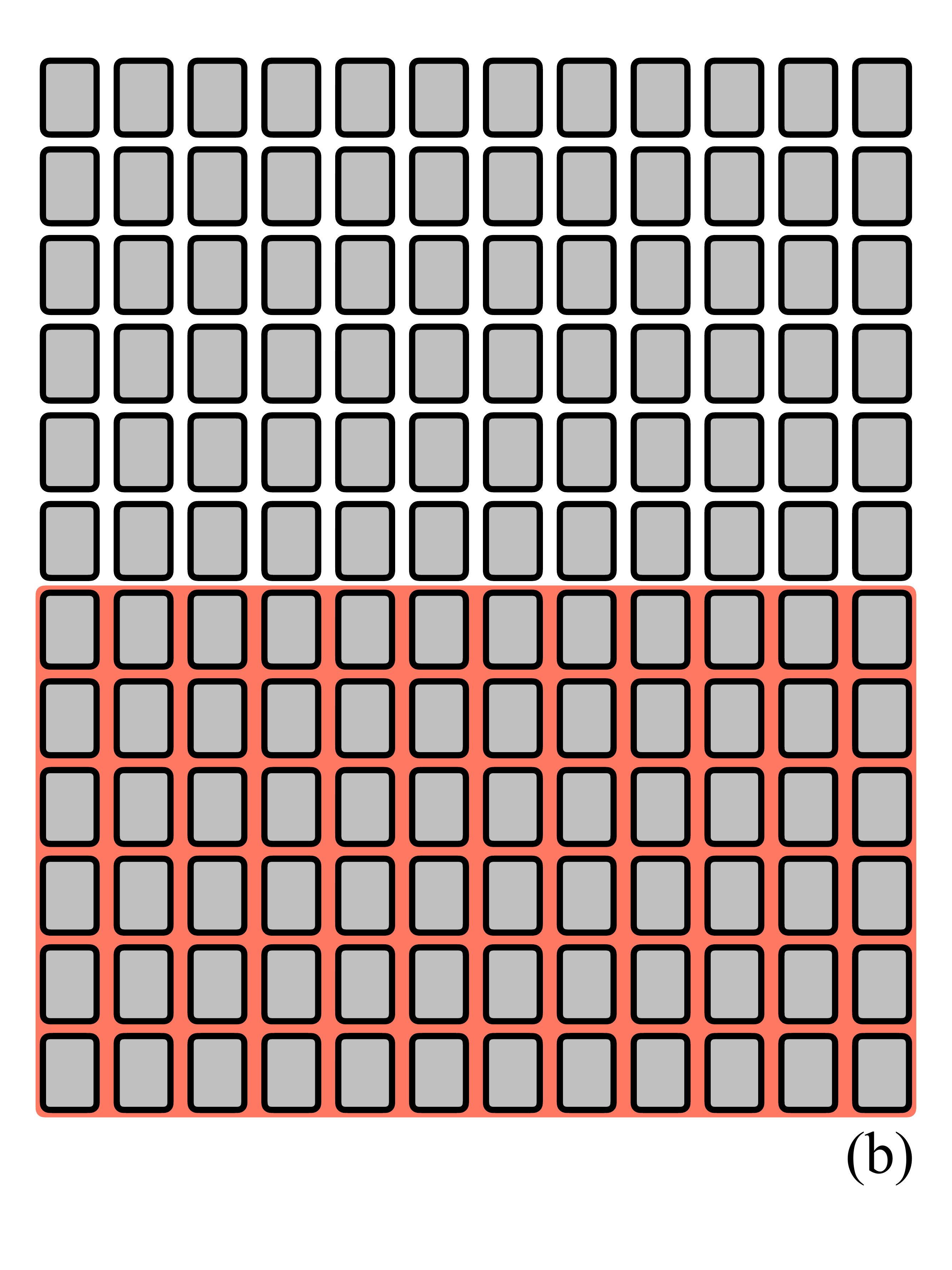}

\vspace*{-1.25em}

\includegraphics[width=0.47\columnwidth]{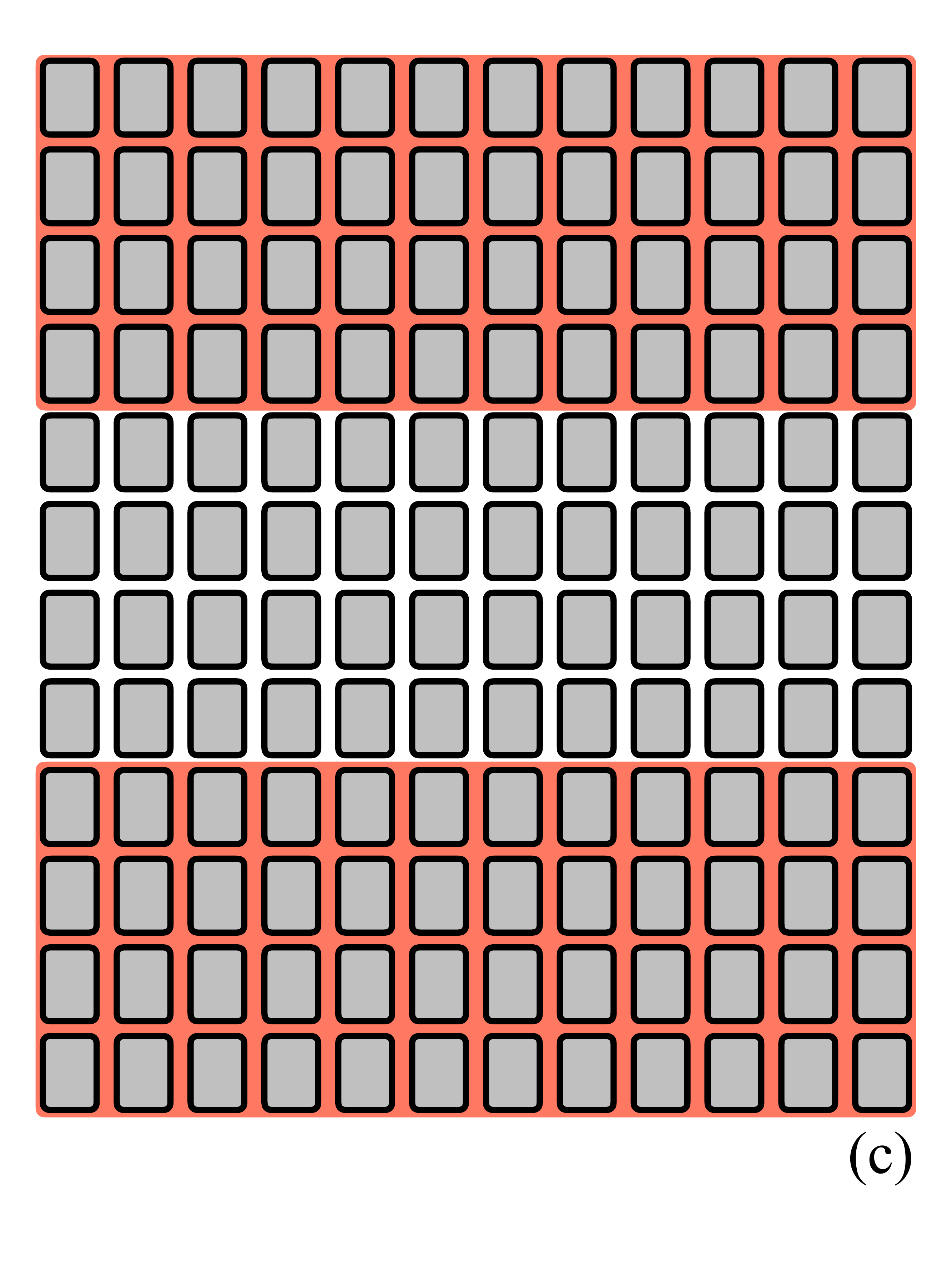}
\hspace*{0.25em}
\includegraphics[width=0.47\columnwidth]{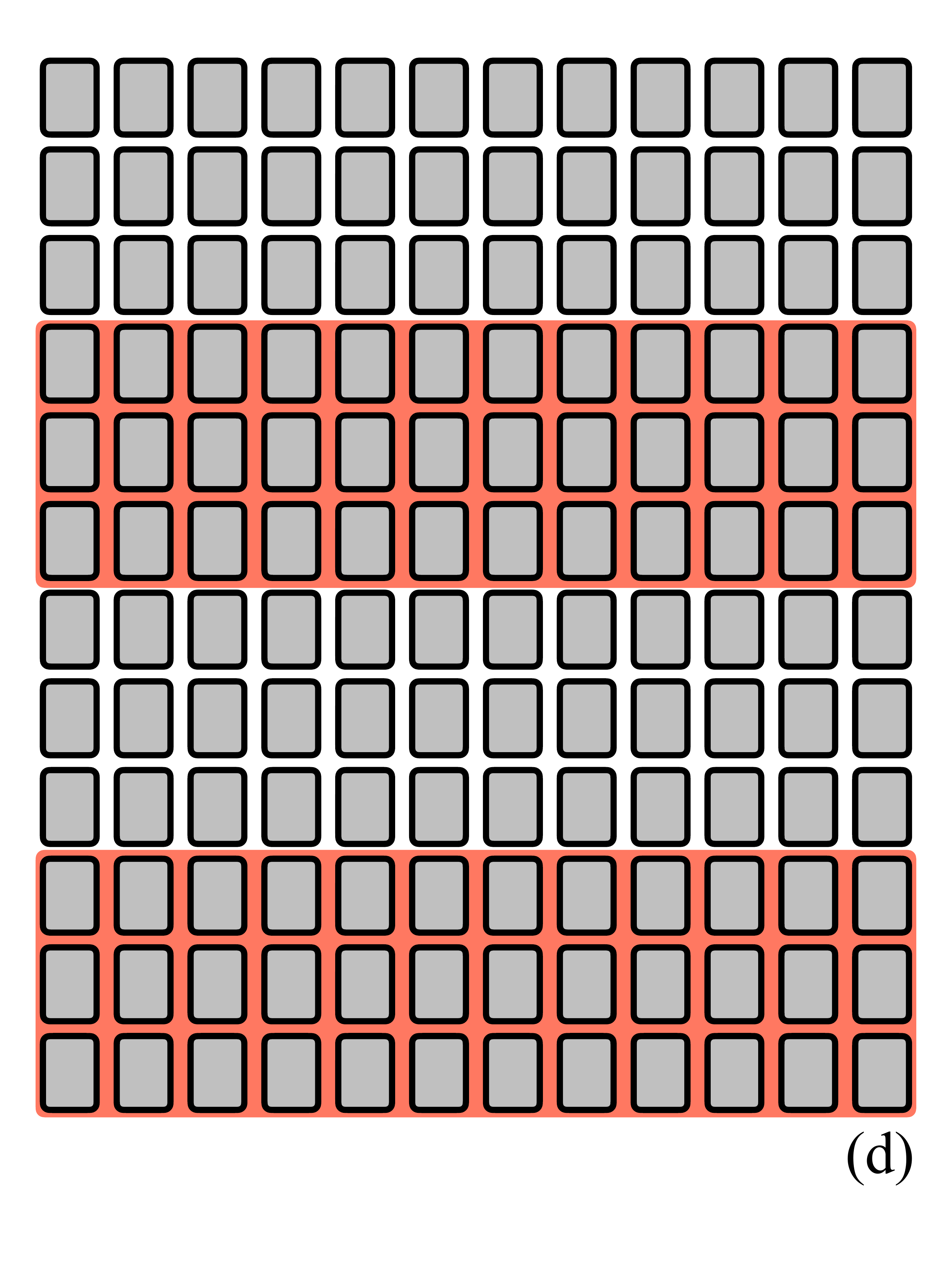}

\vspace*{-1.75em}

\caption{
Sketch of the different patch geometries used on the D-Wave 2X quantum
annealer chip. Each gray block represents a K$_{44}$ call with $8$
sites. (a) A zoom of such cell. (a) -- (d) he shading
represents the different patches used from $M = 1$ (a) to $M = 4$ (d).
}
\label{dw2x}
\end{figure}

Patch-planted instances generated using hard patches are consistently
harder than patch planted instances assembled from easy patches with the
mean value of $\ya$ for both cases being $3.214(5)$ and $2.957(4)$,
respectively. We note that the approach pioneered in
Ref.~\cite{marshall:16} applied to the production of patches could be
combined with patch planting to generate unusually hard planted
problems. It is also interesting to compare the complexity of the
patched instances with random instances. We show the distribution of
$\ya$ for $L=8$ [Figs.~\ref{RH}(a) and \ref{RH}(b)] and $L=12$
[Figs.~\ref{RH}(c) -- \ref{RH}(e)] with different patch sizes and random
instances. One can see that while the patched instances are generally
easier than the random instances, they are not necessarily trivial.
There is clear overlap between the distributions, i.e., by mining the
data one can obtain problems of comparable typical complexity.  Note
also, that the typical complexity grows with increasing patch size for a
fixed system size.

Finally, we comment on the performance of parallel tempering (PT) on
patched instances. Because population annealing and parallel tempering
have a similar performance in both thermal sampling and optimization,
and given that the entropic family size correlates strongly with the
integrated autocorrelation time (characteristic measure of hardness of
parallel tempering) \cite{wang:15e}, it is natural to expect the
proposed patch-planted instances to be hard also for PT. To this end, it
is noteworthy to mention recent results that analyze the performance of
PT with isoenergetic cluster moves (ICM) (see
Ref.~\cite{zhu:16y}) in solving patch-planted instances
\cite{karimi:17}.  PT combined with ICM has been found to be one of the
best classical heuristics in solving hard optimization problems
\cite{mandra:16b}. However, Ref.~\cite{karimi:17} clearly show that PT is
not able to efficiently solve patch-planted instances (see Figs.~8 -- 10).

\subsection{Experiments on the D-Wave quantum annealer}
\label{chimera}

We complement the numerical studies on three-dimensional spin glasses by
experiments on the D-Wave 2X quantum annealer.  For this purpose, we
patch plant problems on the native topology of the machine and measure
the probabilities to find the ground state $p_{\rm{succ}}$ over multiple
runs. In addition, we compare to random problems and show correlation
plots between the success probabilities and $\ya$.

\begin{figure}[htb]
\includegraphics[width=1.0\columnwidth]{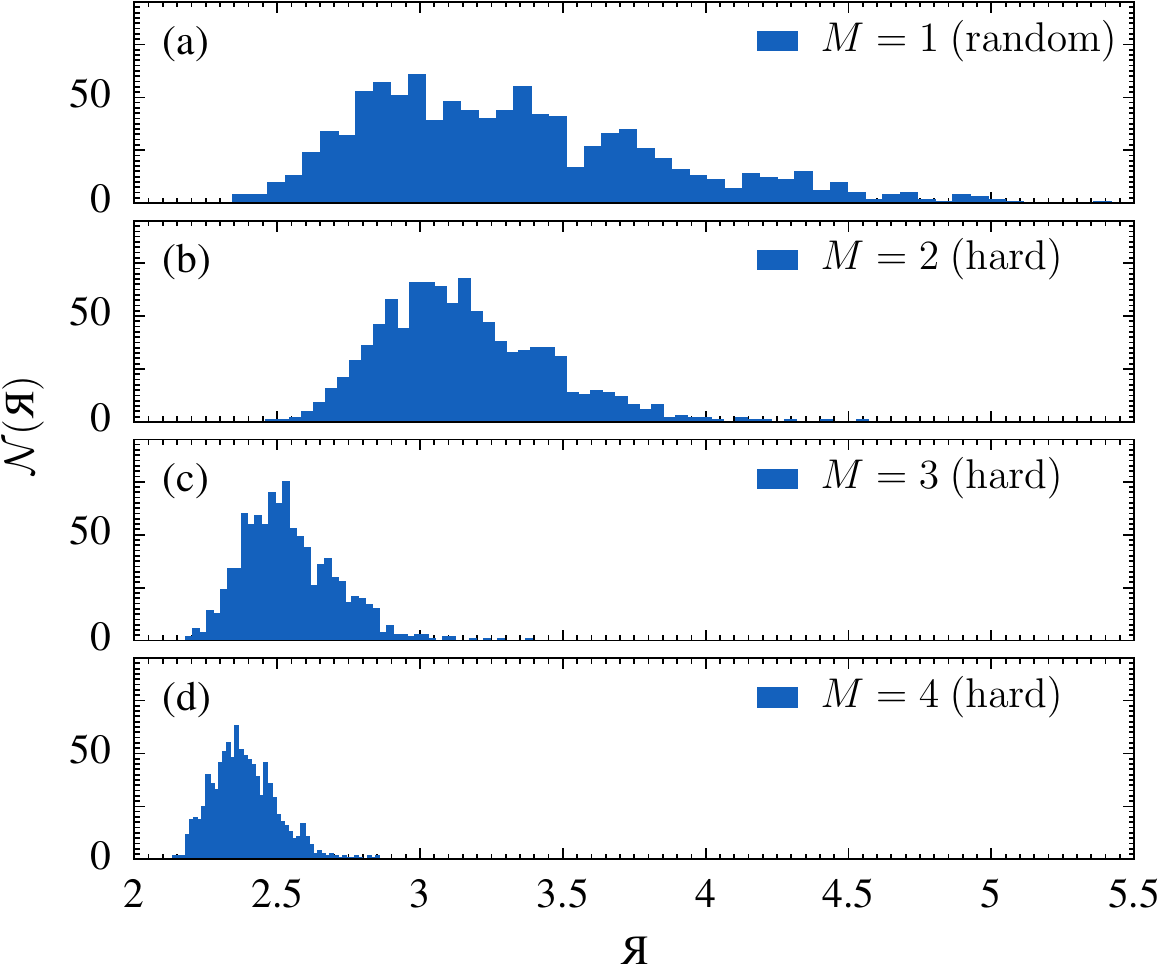}
\caption{
Distributions of $\ya$ on a chimera graph with $N=1097$ for random
instances and patched instances with different number of patches $M$.
There are $1000$ instances each, and the patched instances were chosen
from the hardest ones out of $10^4$ instances in each class. Note that
the patched problems with $M = 2$ (b) and random (a) are
comparable. Panels (c) and (d) show that problems become easier for
smaller patches, i.e., a larger number of patches.}
\label{DWXi}
\end{figure}

\begin{figure}[htb]
\includegraphics[width=1.0\columnwidth]{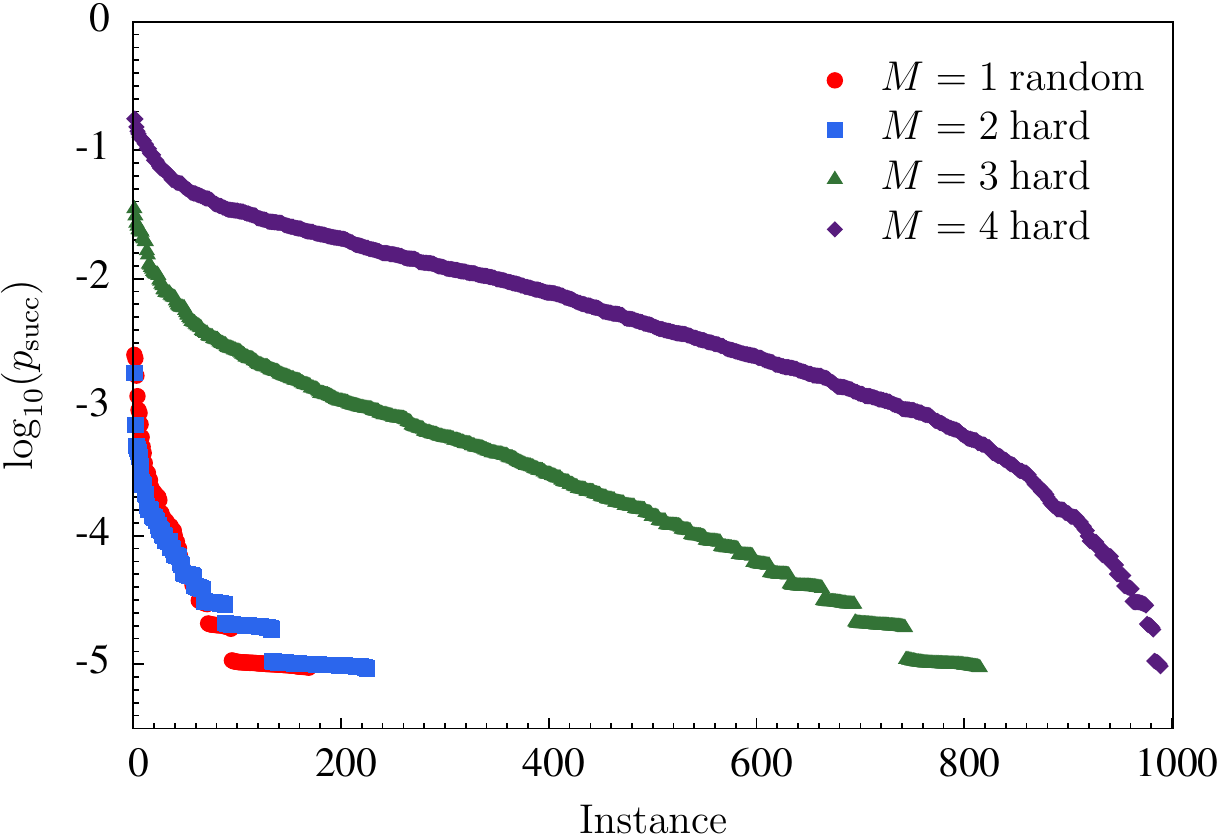}
\caption{
Sorted probabilities to find the ground state $p_{\rm{succ}}$ from
experiments with the D-Wave 2X quantum annealer for $N = 1097$ sites and
different number of patches $M$.  As in Fig.~\ref{DWXi}, random
instances and instances with $M = 2$ are comparable.
}
\label{psucc}
\end{figure}

\begin{figure}[htb]
\includegraphics[width=1.0\columnwidth]{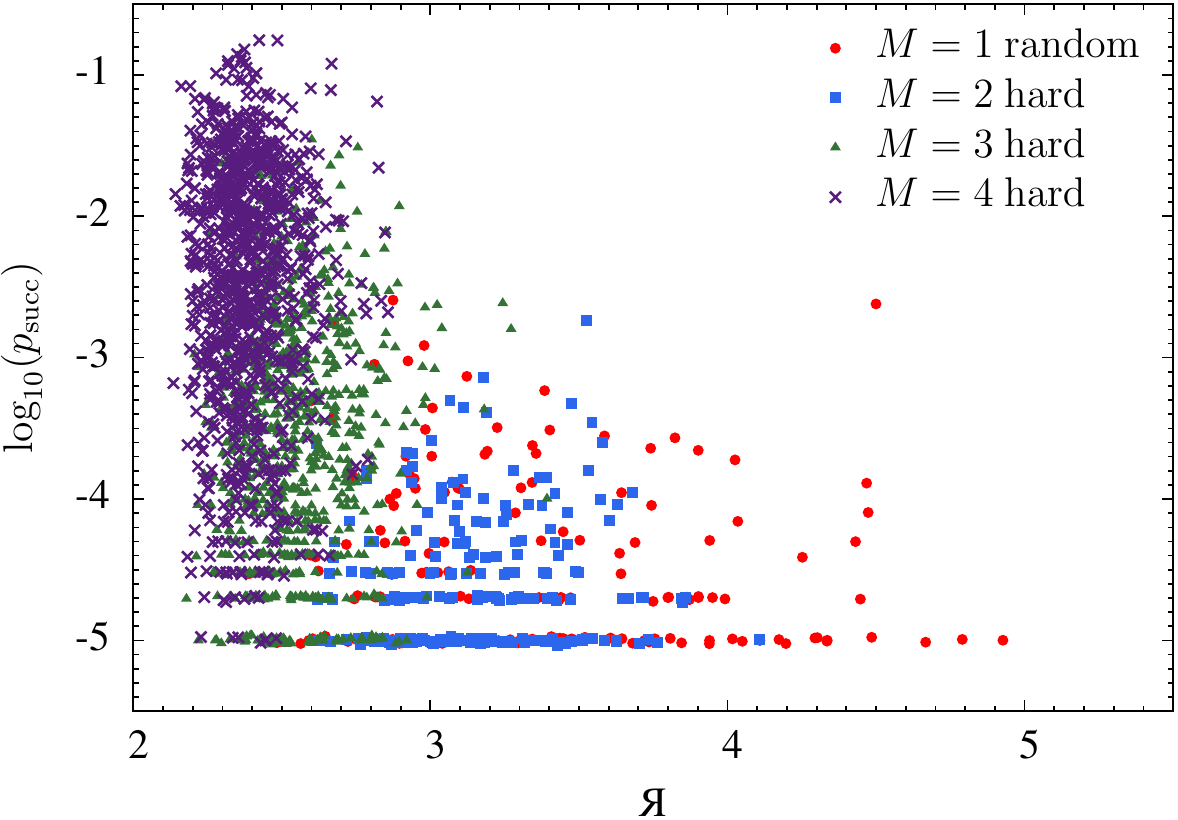}
\caption{
Correlation of $\ya$ and $p_{\rm{succ}}$ for experiments on the D-Wave
machine with $N=1097$. Data for random instances, as well as instances
with different numbers of patches $M$.
}
\label{PXi}
\end{figure}

The topology of the machine with $N = 1097$ working qubits is cut into
two, three, and four patches; see Fig.~\ref{dw2x} for a graphical
representation. For each experiment, we study $10^3$ instances. For the
patch planted instances, we first generate $10^4$ patch planted problems
from random patches and then use $\ya$ to select the $10^3$ hardest
ones.  The distributions of $\ya$ for the problems studied is shown in
Fig.~\ref{DWXi}. One can see that for an increasing number of patches
$M$ the problems become computationally easier. However, again by mining
the data as done above results in hard problems. Figure \ref{psucc}
shows the sorted success probabilities for the $1000$ problems studied
for different number of patches. One can see that problems with $M = 4$
are computationally much easier. It remains to be tested if changing the
shape of the patch could make the problems harder.  For example, the four
patches could be chosen to be comprised of $6 \times 6$ K$_{44}$ cells.
Finally, Fig.~\ref{PXi} shows a correlation plot between success
probabilities and $\ya$. As can be seen, there is a good correlation
between these two quantities, especially for larger patches. Experiments
(not shown) suggest that the correlation becomes more pronounced for
larger system sizes.  With some data mining and only a $10$-fold
overhead, instances with two patches $M = 2$ have approximately the same
complexity as the random ones, which are harder than instances with
three patches and four patches.  Statistics of the success probabilities
are shown in Table~\ref{statistics}.

\begin{table}
\caption{
Statistics of the D-Wave 2X quantum annealer success probability
$p_{\rm{succ}}$ in Fig.~\ref{psucc} for random instances and patched
instances with different number of patches $M$.  There are $10^3$
instances each, and the patched instances were chosen from the
hardest ones out of $10^4$ instances in each class.  $p_{\rm{min}}$,
$p_{\rm{max}}$, and $p_{\rm{ave}}$ are the minimum, maximum and average
values of $p_{\rm{succ}}$, respectively and $f$ is the fraction of
instances with $p_{\rm{succ}}=0$.
\label{statistics}
}
{\scriptsize
\begin{tabular*}{\columnwidth}{@{\extracolsep{\fill}} l l l l r}
\hline
\hline
               &Random      & $M = 2$     & $M = 3$   & $M = 4$ \\
\hline
$p_{\rm{min}}$ & $0$     & $0$      & $0$    & $0$ \\
$p_{\rm{max}}$ &$0.00240(32)$   & $0.00137(59)$   & $0.0326(32)$  & $0.173(9)$ \\
$p_{\rm{ave}}$ &$0.0000204(46)$ & $0.0000129(24)$ & $0.00115(10)$ & $0.0127(7)$ \\
$f$            &$0.831(12)$     & $0.775(13)$     & $0.185(12)$ & $0.011(3)$ \\
\hline
\hline
\end{tabular*}
}
\end{table}

\section{Summary}
\label{sec:summary}

We have introduced the concept of patch planting to create planted
solutions to Ising-type optimization problems for arbitrarily large
systems. The method does not restrict the values of the couplers and
works for any topology that can be decomposed into patches.  We studied
in detail the scaling of the typical complexity of the patched
instances and compared this to random instances using population annealing
Monte Carlo and the D-Wave 2X machine.  From our results it is clear
that one should use as large patches as possible to more faithfully
reproduce the hardness of random problems.  Patch planting is easy to
implement and could be used to generate benchmark instances for future
generations of quantum devices, as well as classical algorithms and any
other novel hardware. The approach is generic in that solutions could
also be planted for other paradigmatic optimization problems (e.g., the
traveling salesman problem) with only minor modifications.

\begin{acknowledgments}

We thank Firas Hamze, Jon Machta, and Ethan Brown for helpful
discussions. H.G.K.~would like to thank I.~P.~Freely and I.~M.~A.~Wiener
for inspiration at the early stages of the project.  W.W.~and
H.G.K.~acknowledge support from the National Science Foundation (Grant
No.~DMR-1151387). The research of H.G.K.~is based upon work supported in
part by the Office of the Director of National Intelligence (ODNI),
Intelligence Advanced Research Projects Activity (IARPA), via MIT
Lincoln Laboratory Air Force Contract No.~FA8721-05-C-0002. The views
and conclusions contained herein are those of the authors and should not
be interpreted as necessarily representing the official policies or
endorsements, either expressed or implied, of ODNI, IARPA, or the
U.S.~Government.  The U.S.~Government is authorized to reproduce and
distribute reprints for governmental purpose notwithstanding any
copyright annotation thereon.  We thank Texas A\&M University for access
to their Ada and Curie clusters.

\end{acknowledgments}

\bibliographystyle{apsrevtitle}
\bibliography{refs,comments}

\begin{thebibliography}{48}
\bibitem[{com({\natexlab{b}})}]{comment:alan}
\bibinfo{note}{{The idea of patch planting is inspired by patchwork approaches
  to compute ground states of Ising spin glasses \cite{thomas:08}. However, in
  this case the goal is to plant a solution and not find one.}}
\end{thebibliography}

\end{document}